\documentclass[preprint,prl,showpacs,floatfix]{revtex4}

\usepackage{graphicx}
\usepackage{amsmath}
\usepackage{nameref,hyperref}

\begin{document}

\title
{
Simple Elastic Network Models for Exhaustive Analysis of Long Double-Stranded DNA Dynamics with Sequence Geometry Dependence
}

\author
{
Shuhei Isami$^1$, Naoaki Sakamoto$^{1,2}$, Hiraku Nishimori$^{1,2}$,  Akinori Awazu$^{1,2}$
}

\affiliation
{
$^1$Department of Mathematical and Life Sciences, Hiroshima University,
$^2$Research Center for Mathematics on Chromatin Live Dynamics. \\
Kagami-yama 1-3-1, Higashi-Hiroshima 739-8526, Japan.
}


\begin{abstract}
Simple elastic network models of DNA were developed to reveal the structure-dynamics relationships for several nucleotide sequences. First, we propose a simple all-atom elastic network model of DNA that can explain the profiles of temperature factors for several crystal structures of DNA. Second, we propose a coarse-grained elastic network model of DNA, where each nucleotide is described only by one node. This model could effectively reproduce the detailed dynamics obtained with the all-atom elastic network model according to the sequence-dependent geometry. Through normal-mode analysis for the coarse-grained elastic network model, we exhaustively analyzed the dynamic features of a large number of long DNA sequences, approximately $\sim 150$ bp in length. These analyses revealed positive correlations between the nucleosome-forming abilities and the inter-strand fluctuation strength of double-stranded DNA for several DNA sequences.
\end{abstract}


\maketitle

\section{Introduction}
Elastic network models of proteins, including all-atom models\cite{EN1,EN2} and coarse-grained models\cite{EN3,EN4,EN9,EN10,EN11,EN12,EN13} represent some of the simplest and most powerful types of theoretical models that can accurately reveal structure-dynamics relationships and the mechanisms underlying a protein's functional activities \cite{EN5,EN6,EN7,EN8,EN14}. Such models have also been widely employed to accurately reproduce the temperature factors on the crystal structure of a protein via normal-mode analysis. These models can thus demonstrate the large and slow deformations of proteins that are essential for protein functions, but remain difficult to demonstrate via all-atom molecular dynamics simulations.

 Along with proteins, DNA is the most important biomolecule for the activities of all living organisms. Recent developments in molecular biology have revealed that DNA contains several functional regions. Therefore DNA is no longer considered to have the sole function in the storage of genetic information, but is now known to be actively involved in gene regulation\cite{cell,gene0,gene3,gene4}, insulator activity\cite{ins1,ins2,ins3,ins4,ins5,ins6}, and in the construction of chromosomal architectures\cite{gene1,gene2,chro1,chro2} such as heterochromatins and topologically associated domains through nucleosome formation and protein bindings\cite{tad1,tad2,tad3,tad4,tad5}. The functional behavior of each strand of DNA is determined not only by chemical aspects of the nucleotides and base pairs but also by its physical characteristics such as the structure and dynamics of the nucleotide sequences in each functional region. However, comprehensive understanding of the physical properties of nucleotide sequences lags far behind the knowledge accumulated of their biochemical aspects \cite{cell,gene0,gene1}.

Since the last century, much progress has been made in revealing the physical aspects of DNA using all-atom normal-mode analysis\cite{full1,full2} and molecular dynamics simulations\cite{full3,full4,full5,cg1,full6,full7,full8}. Several coarse-grained models of DNA (and RNA) have also been proposed. Some of these describe the detailed shape of each nucleotide using three or more particles\cite{cg0,cg30,cg3,cg4,cg40,cg6,cg9}, whereas others describe each base pair by simply one or two particles\cite{cg2,cg5,cg50,cg7,cg8}. Molecular dynamics simulations and normal-mode analysis of these models have identified the flexibilities, nucleosome-forming abilities, and zip-unzip transitions of the double helices of some specific DNA sequences from tens to a few hundred base pairs in length. Although these methods have proven to be very powerful for the analysis of the physical aspects of DNA, molecular dynamics simulations are not suitable for conducting exhaustive analyses of several sequences simultaneously owing to the high computational costs of such extensive simulations; for example, analyses of whole genomic and whole possible sequences. Moreover, the normal-mode analysis of all-atomic models of long DNA sequences is also computationally heavy.

Alternatively, data-driven methods have been proposed for determining the mechanical properties of DNA with respect to the helical parameters and local flexibilities of base pairs from X-ray crystal structure analysis and all-atom molecular dynamics \cite{nuc1,nuc2,nuc3,nuc4,nuc5,nuc6}. These methods also seem to be powerful and can be applied to the analysis of the mechanical properties of several DNA sequences simultaneously. However, the methods thus far proposed have focused on static and local mechanical properties. Therefore, they are not particularly useful for the study of the functional contribution of the dynamic and correlated motions of DNA.

The objective of this study was to construct a simple coarse-grained elastic network model of DNA to allow for exhaustive analysis of the dynamic correlated motions of several long DNA sequences. For this purpose, we first constructed a simple all-atom elastic network model of short DNA sequences based on the method introduced by Tirion\cite{EN1} for the modeling of protein dynamics. We confirmed that this model could accurately reproduce the temperature factors of some DNA fragments from data obtained with X-ray crystal structure analysis.

Second, we developed a simple coarse-grained elastic network model of long DNA sequences, where each nucleotide is described by only one node. We confirmed that this simplified model has lower computational costs but can nonetheless reproduce the nucleotide sequence-dependent dynamics revealed by the all-atom elastic network model.

Finally, through the normal-mode analysis of this coarse-grained model, we conducted an exhaustive analysis of the general features of the sequence-dependent structure-dynamics relationships among several DNA sequences. We specifically focused on the dynamic properties of a large number of long DNA sequences, approximately $\sim 150$ bp in length (with respect to the length of the nucleosome-forming regions), for the genomes of some model organisms as well as for random sequences with several A, T, C, G ratios. Through these analyses, we found that the dynamic aspects of DNA are highly influenced by their sequences, and found positive correlations between the nucleosome-forming abilities and inter-strand fluctuations of double-stranded DNA.

In particular, we focus on the geometry dependencies of the dynamics of several DNA sequences, since a recent study demonstrated that the geometry of DNA sequences has a dominant contribution to their mechanical features\cite{cg40}. In the following arguments, for simplicity, we use ``sequence-dependent'' to mean ``dynamics that depend on sequence geometry''.

\section{Models and Methods}
\subsection{Basic structures of Elastic Network Models of DNA }
In order to construct elastic network models of DNA, the basic DNA structure first needs to be determined. In the following arguments, we construct two types of models: i) a simple all-atom elastic network model (AAENM) to reproduce the characteristics of the crystal structures of DNA, and ii) a simple coarse-grain elastic network model (CGENM) to reproduce the characteristics of the AAENM. We obtained the basic DNA structures in the following two ways for the respective purposes of constructing each model.

For construction of the AAENM, we employed the atom coordinate sets of several naked DNA crystal structures included in the Protein Data Bank (PDB) as the given basic structures of the model. We used $9$ DNA crystal structures containing only DNA and water molecules under different conditions of crystallization, where all temperature factors are given as positive values (Table 1). The suitability of the model was evaluated by comparisons of the temperature factors obtained between the model and those obtained from X-ray crystal structure analysis.

The objective of our constructed arguments was to unveil the sequence-dependent dynamic features of several long DNA sequences simultaneously. In recent crystal structure analysis, only shorter DNA sequences (i.e., less than $12$ bp in length) were studied. On the other hand, several types of helical parameter sets, base pair parameters, and base step parameter sets have been proposed through experiments or molecular dynamics simulations\cite{full3,heri1,heri2,heri3,heri4,heri5,heri6}(Table 2, \nameref{S1_Table}). Here, we used the application X3DNA\cite{3DNA} to obtain the coordinates of each atom for any sequence from these helical parameters, and used these parameters to construct the AAENMs and CGENMs of longer DNA sequences (for detailed methods of the generation of the atom coordinates, see Lu and Olson \cite{3DNA}). In the following arguments, we compare the characteristics of the AAENM and CGENM constructed by X3DNA from three different helical parameter sets: i) a parameter set obtained by an {\it in vitro} experiment and X-ray crystal structure analysis\cite{cg6,heri1,heri2,heri3} (Table 2), ii) a parameter set obtained only by the X-ray crystal structure analysis, and iii) a parameter set obtained by all-atom molecular dynamics simulations\cite{nuc4,heri5,heri6} (\nameref{S1_Table}).

\begin{table}
\begin{center}
\includegraphics[width=12.0cm]{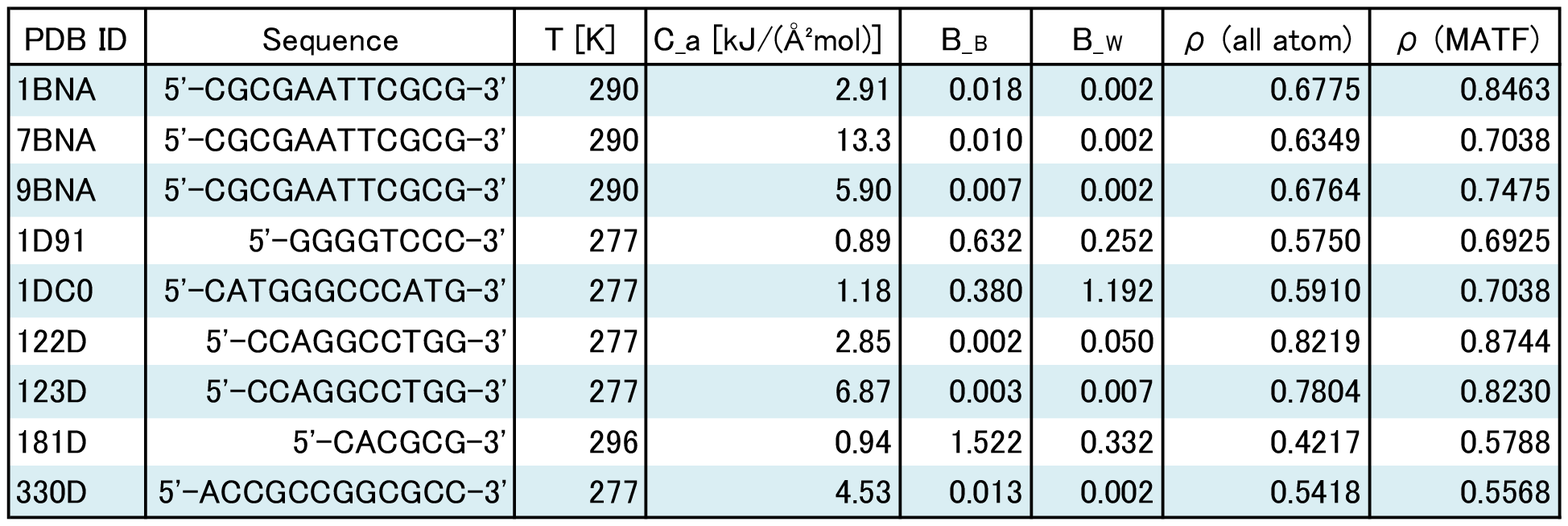}
\end{center}
\caption{{\bf Information of the analyzed DNA segments.} PDB ID, sequences of X-ray crystal structure analysis of DNA fragments, parameter sets of the AAENM for each DNA structure, and correlation coefficients of $TF_i$ and $MATF_i$ between the AAENM and X-ray crystal structure.}
\end{table}

\begin{table}
\begin{center}
\includegraphics[width=12.0cm]{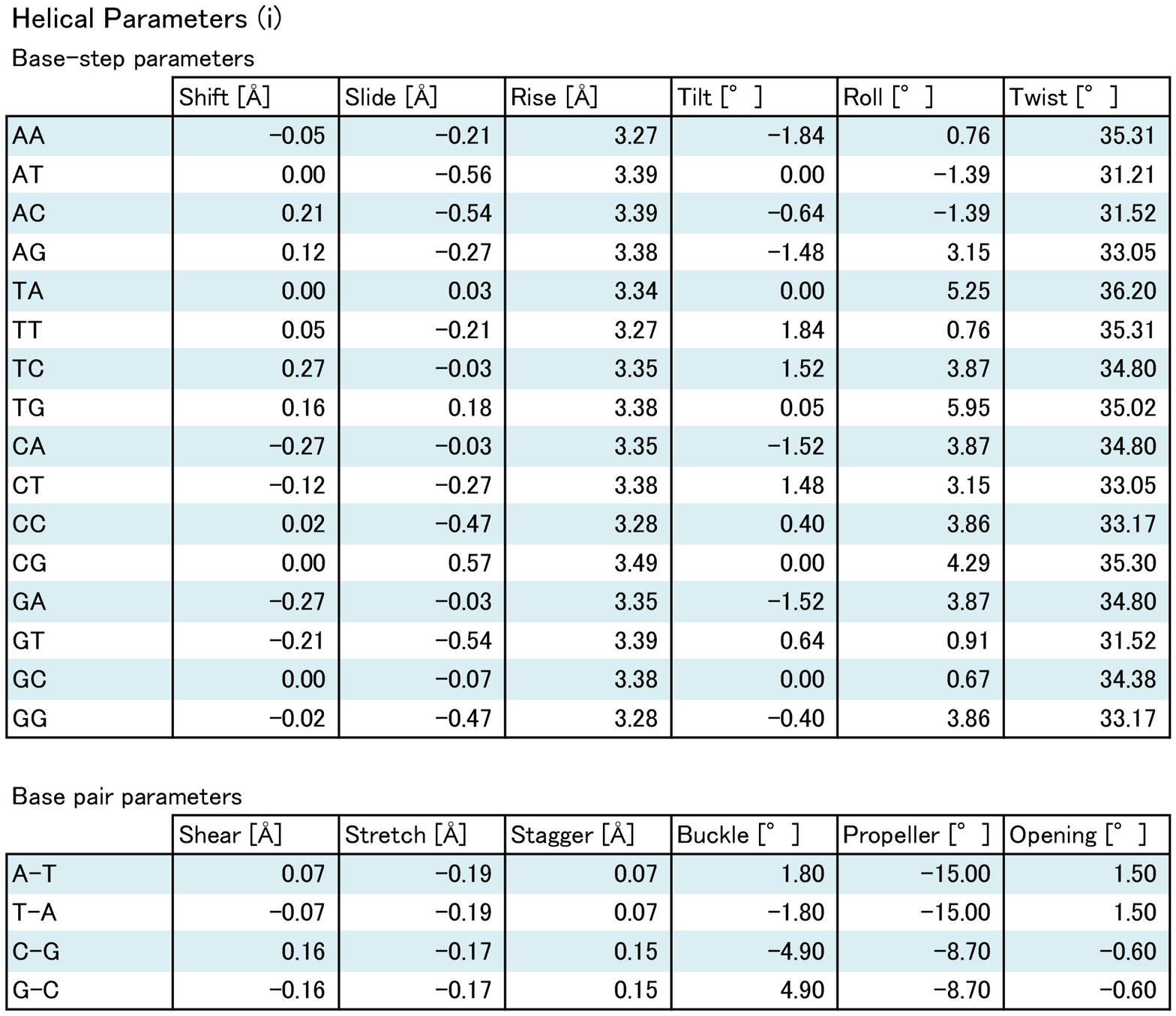}
\end{center}
\caption{{\bf Helical parameter sets.} Helical parameter sets (i) obtained by {\it in vitro} experiments and X-ray crystal structure analysis\cite{cg6,heri1,heri2,heri3}.}
\end{table}

\subsection{All-atom Elastic Network Model of DNA}
A simple all-atom elastic network model of double-stranded DNA was constructed based on the model proposed by Tirion\cite{EN1}. In this model, we regard all of the atoms of given DNA sequences as the nodes comprising the elastic network. For the analysis of the crystal structures of DNA involving water molecules, we also regard the oxygen atoms of water around DNA as the nodes of the elastic network. We define the mass and position of atom $i$ as $m_i$ and ${\bf r}_i$ (${\bf r}_i = (x_i, y_i, z_i)$), respectively. The potential $V$ of all atoms is given as
\begin{equation}
V = \sum_{i,j} \frac{C_a}{2}(|{\bf r}_i - {\bf r}_j| - |{\bf r}_i^0 - {\bf r}_j^0|)^2\theta(R_{i}+R_{j}+R_c-|{\bf r}_i^0 - {\bf r}_j^0|) + \sum_{boundary} \frac{B_iC_a}{2}({\bf r}_i - {\bf r}_i^0)^2 .
\end{equation}

Here, ${\bf r}_i^0$ is the position of atom $i$ of the basic DNA structure, as described above.

The first term indicates the interaction potential among atoms that are spatially closed in the basic DNA structure (Fig. 1(a)). Here, $R_{i}$ refers to the Van der Waals radius of atom $i$, $R_c$ is an arbitrary cut-off parameter that models the decay of interactions with distance, and $\theta$ indicates the Heaviside function, where $\theta(z) = 1$ ($\theta(z)=0$) for $z \ge 0$ ($z < 0$).
We assume $R_c = 2.0\AA$, which is empirically considered to be an appropriate range for biomolecules, at least for proteins\cite{EN1}. The results of the arguments presented in this paper were qualitatively unchanged for the appropriate range of $R_c$. In the crystal structures of DNA, the motions of water molecules are also restricted by the crystal packing. Thus, for all atoms, we assume that spatially closed pairs of atoms are connected by linear springs with their respective natural lengths. The elastic coefficient $C_a$ ($[kJ/(\AA^2mol)]$) is a phenomenological constant, which is assumed to be the same for all interacting pairs.

\begin{figure}
\begin{center}
\includegraphics[width=8.0cm]{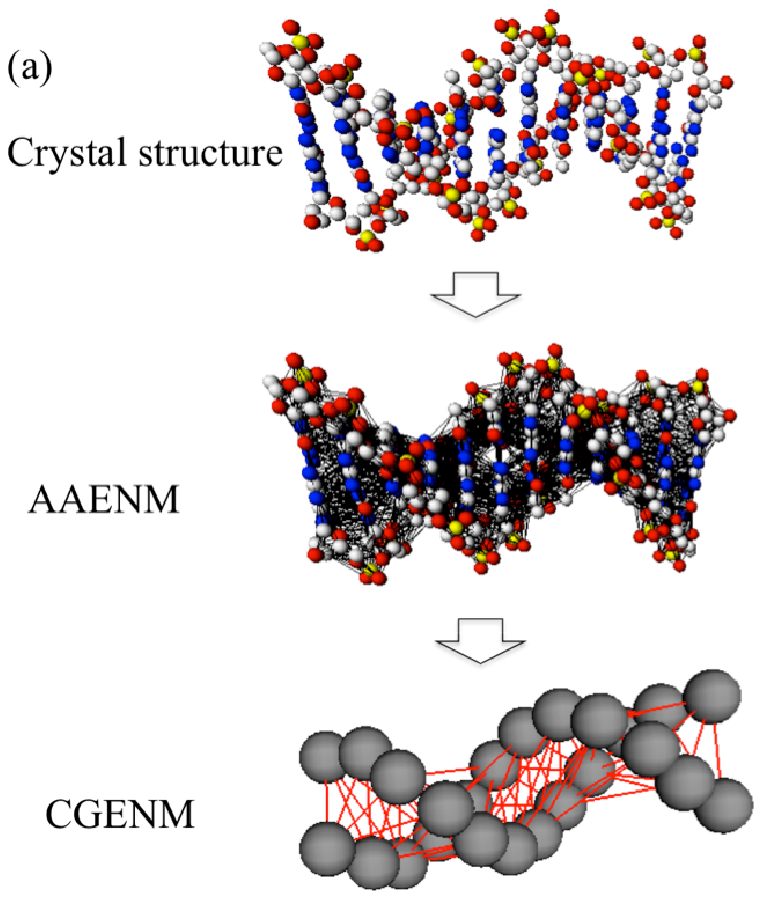}
\includegraphics[width=8.0cm]{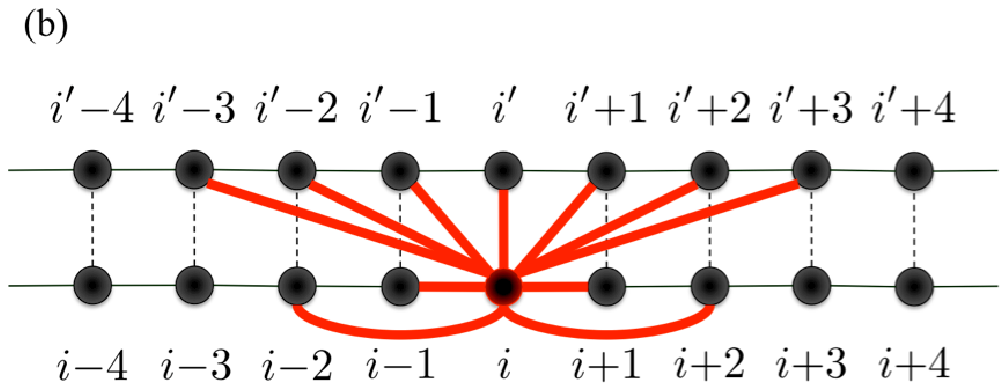}
\includegraphics[width=8.0cm]{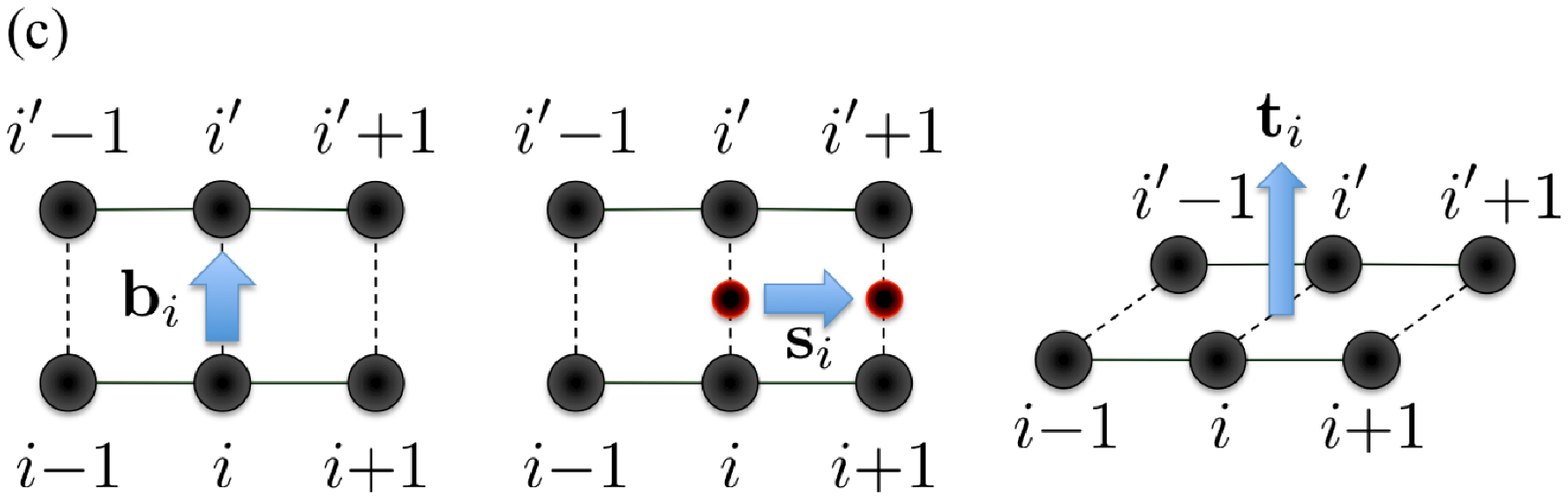}
\end{center}
\caption{Illustrations of (a) all-atom elastic network models (AAENMs) from the crystal structures, and the coarse-grained elastic network models (CGENMs) from the AAENMs; (b) detailed interactions among the nucleotides (nodes) of the CGENM where nucleotide $i$ interacts with all nucleotides connected by the 11 bold curves; and (c) ${\bf b}_i, {\bf s}_i$, and ${\bf t}_i$. }
\end{figure}

The second term indicates the boundary effects of each DNA and water molecule in each DNA crystal structure. This term plays a crucial role for the analysis of the fluctuations of the crystal structure of DNA, such as temperature factors, since the fluctuations of the nucleotides at the edges of DNA and water molecules are restricted due to the following facts.

In the crystal of DNA, the motion of edges at upper and lower streams of each DNA segment (left and right edges in Fig. 1(a)) is influenced by atoms of other adjacent segments in the long-axis direction of DNA segments. Moreover, the motions of water molecules around each DNA segment are influenced by atoms of other adjacent water molecules or DNA segments in the direction perpendicular to the longitudinal axis of the DNA segment. Thus, we need to consider the second term of Eq. (1) to implement such effects, where $B_{j}$ indicates the strength of such effects for atoms in the edge nucleotides of each DNA and water molecule, respectively.

Remarkably, as shown in the Results section, the distributions of the temperature factor of atoms exhibited various patterns from the same sequence of crystallized DNA (\nameref{S1_Fig} and \nameref{S2_Fig}) owing to the dependency on the conditions of crystallization. Therefore, in order to compare the characteristics of the present DNA model to those of the crystal structure of DNA, appropriate values of $B_j$ need to be assigned to the atom $j$ that belongs to the edge base pairs and water molecules. For simplicity, we assign $B_j = B_B$ and $B_w$ to the atoms $j$ at the edge base pairs and the water molecules, respectively, and $B_j = 0$ otherwise.

It is noted that the internal structure of the crystal of DNA is spatially anisotropic. Thus, it is reasonable to assume that the interactions among local parts of the crystal of DNA exhibit different strengths in different directions. Accordingly, in general, the strengths of the restrictions of atoms belonging to the edge of DNA differ from those of water molecules. Thus, we assume that $B_{B}$ and $B_{w}$ are different values (Table 1).

For simplicity, the mass of each atom is assumed as a constant value ($m_i = m = 10 \time 10^{-3} / N_A [kg]$, $N_A = 6.02214129 \times 10^{23} [ /mol]$ is Avogadro's number). However, we confirmed that the results were almost identical when using the precise masses of the atoms.

\subsection{Coarse-grained Elastic Network Model of DNA}
A simple coarse-grained elastic network model of double-stranded DNA, where each nucleotide is described as one node, was constructed as follows (Fig. 1(a)). We define the coordinate of the C1' carbon of nucleotide $i$, ${\bf x}_i$ (${\bf x}_i = (x_i, y_i, z_i)$), as the position of nucleotide $i$, and regard the motion of the C1' carbon as that of the nucleotide. Here, we assume that the mass of the C1' carbon obeys $m_i = 10 \time 10^{-3} / N_A [kg]$. The potential $V$ of all nucleotides is given as
\begin{equation}
V = \sum_{i,j} \frac{C_g}{2}(|{\bf x}_i - {\bf x}_j| - |{\bf x}_i^0 - {\bf x}_j^0|)^2.
\end{equation}
Here, ${\bf x}_i^0$ is the position of nucleotide $i$ of the basic structure of DNA, as defined above. We assume that nucleotides $i$ and $i'$ belong to the same base pair. For nucleotide $i$, the sum is restricted to the pair of nucleotides in the same base pair ($j = i'$), in the neighboring base pairs ($j = i+1, i'+1, i-1, i'-1$), in the next neighboring bases pairs ($j = i+2, i'+2, i-2, i'-2$), and in the next-next neighboring bases in another strand ($j = i'+3, i'-3$) (Fig. 1(b)). The elastic coefficient $C_g$ ($[kJ/(\AA^2mol)]$) is a phenomenological constant that is assumed to be the same for all interacting pairs. This model is considered as a simplified version of previously proposed one-site-per-nucleotide models\cite{cg2,cg7}.

\subsection{Normal-mode Analysis}
An overview of the theory of normal-mode analysis is provided in several recent studies\cite{EN1,EN2,EN3,EN4,EN5,EN6,EN7,EN8,EN9,EN10,EN11}. Thus, we here briefly show the results of this analysis. For this analysis, we define ${\bf q}(t)$ (${\bf q}=({\bf q}_1, {\bf q}_2, ... {\bf q}_N)$, ${\bf q}_i = (x_i, y_i, z_i)$) as a $3N$-dimensional position vector, and ${\bf q}^0$ as the position vector of the basic structure. Here, ${\bf q} = ({\bf r}_1, {\bf r}_2, ... {\bf r}_N)$ for the AAENM and ${\bf q} = ({\bf x}_1, {\bf x}_2, ... {\bf x}_N)$ for the CGENM. The motions of small deviations of ${\bf q}(t)$ from ${\bf q}^0$, ${\bf \delta q}(t) = {\bf q} - {\bf q}^0$ obey
\begin{equation}
{\bf \delta q}(t) = \sum_{\omega_k \ne 0} A_k{\bf v}_ke^{i \omega_k t}
\end{equation}
where $-(\omega_k)^2$ and ${\bf v}^k = ({\bf v}^k_1, {\bf v}^k_2, ... , {\bf v}^k_N)$ (${\bf v}^k_i = (v^k_{x_i}, v^k_{y_i}, v^k_{z_i})$) are the $k$-th largest eigenvalue and its eigenvector of the $3N \times 3N$ Hessian matrix ${\bf H}$ as
\begin{equation}
H_{ij} = -\left(\frac{\partial^2V}{\partial q_i \partial q_j} \right)_{{\bf q}={\bf q}^0}.
\end{equation}

We assume that the system is in thermodynamic equilibrium with temperature $T$. Thus, the amplitude $A_k$ is given as
\begin{equation}
(A_k)^2 = \frac{k_B T}{(\omega_k)^2 m}
\end{equation}
with Boltzmann constant $k_B = 1.3806488 \times 10^{-23} [m^2 kg / s^2 K]$. Using this solution, the mean square fluctuation of the $i$-th atom in the AAENM (${\bf \delta q}_i = {\bf \delta r}_i$) is obtained as
\begin{equation}
AF_i = <|{\bf \delta r}_i|^2> = \sum_{\omega_k \ne 0} \frac{k_B T}{(\omega_k)^2 m}|{\bf v}^k_i|^2
\end{equation}
with Boltzmann constant,
 and the temperature factor is displayed as $TF_i = \displaystyle \frac{8}{3}\pi^2 AF_i$. Here, $<...>$ represents the temporal average.

For the CGENM, we define the mean square fluctuation of the $n$-th nucleotide (${\bf \delta q}_n = {\bf \delta x}_n$) as $CF_n = <|{\bf \delta x}_{n}|^2>$. To consider the motion of the $n$-th nucleotide in the AAENM, we define the average nucleotide motions as ${\bf \delta R}_n = <{\bf \delta r}_j>_{n-th\,\,nucleotide}$. Using this vector, we define the mean square fluctuation of the $n$-th nucleotide as $NF_i = <|{\bf \delta R}_{n}|^2>$. For the motif $mo$ of the $n$-th nucleotide ($mo=$ sugar, base, and phosphoric acid), the average temperature factor of the motif ($MATF_{(mo\,\,in\,\,n-th\,\,nucleotide)}$) is defined as the average of $TF_j$ belonging to each motif of each nucleotide, $<TF_j>_{(mo\,\,in\,\,n-th\,\,nucleotide)} = \displaystyle \frac{8}{3}\pi^2 <AF_j>_{(mo\,\,in\,\,n-th\,\,nucleotide)}$.

\subsection{Treatment of the Temperature Factor in X-ray crystal Structure Analysis}
In order to evaluate the validity of the AAENM, we measured the correlation coefficient between the profile of the temperature factor obtained from PDB data (via X-ray crystal structure analysis) and that obtained from the AAENM based on this crystal structure. It is noted that the temperature factor profiles for some of the PDB data often include unnaturally large or small values. Thus, the correlation coefficients were estimated using data excluding such outliers. In the present evaluations, the value $g_i$ was considered as an outlier if $|g_i-\mu| > s\sigma$, where $\mu$ and $\sigma$ are the average and standard deviation of $\{g_i\}$, respectively, and $s = 2.5$ is used based on the standard arguments of statistics.

\subsection{Evaluations of Anisotropic Fluctuations of DNA}
We also focused on the relationships between the fluctuations of each nucleotide in the AAENMs and CGENMs in the directions parallel to the base pair axis, parallel to the helix axis, and vertical to both the base pair and helix axes.

Here, we name the nucleotides in one and the other strand constructing the $i$-th base pair as the $i$-th and $i'$-th nucleotide. We define the position vectors of the C1' atoms belonging to the $i$-th and $i'$-th nucleotides as ${\bf c}_i$ and ${\bf c}_{i'}$, and consider
\begin{equation}
{\bf b}_i = \frac{{\bf c}_i^0 - {\bf c}_{i'}^0}{|{\bf c}_i^0 - {\bf c}_{i'}^0|},
\end{equation}
\begin{equation}
 {\bf s}_i = \frac{({\bf c}_{i+1}^0 + {\bf c}_{i'+1}^0)-({\bf c}_{i}^0 + {\bf c}_{i'}^0)}{|({\bf c}_{i+1}^0 + {\bf c}_{i'+1}^0)-({\bf c}_{i}^0 + {\bf c}_{i'}^0)|},
\end{equation}
and
\begin{equation}
{\bf t}_i = \frac{{\bf b}_i \times {\bf s}_i}{|{\bf b}_i \times {\bf s}_i|}
\end{equation}
(Fig. 1(c)). It is noted that ${\bf b}_i$ and ${\bf s}_i$ are not orthogonal in general; however, we confirmed that the angles of these vectors were always sufficiently close to $\pi/2$ rad for each $i$.

Using these vectors, the mean square fluctuations of the $i$-th and $i'$-th nucleotides of the AAENM in the directions parallel to the base pair axis are defined as $NF^b_i=<|{\bf \delta R}_{i}{\bf b}_i|^2>$ and $NF^b_{i'}=<|{\bf \delta R}_{i'}{\bf b}_{i}|^2>$, those parallel to the helix axis are defined as $NF^s_{i}=<|{\bf \delta R}_{i}{\bf s}_{i}|^2>$ and $NF^s_{i'}=<|{\bf \delta R}_{i'}{\bf s}_i|^2>$, and those in the torsional direction are defined as $NF^t_{i}=<|{\bf \delta R}_{i}{\bf t}_i|^2>$ and $NF^t_{i'}=<|{\bf \delta R}_{i'}{\bf t}_i|^2>$, respectively.  The mean square fluctuations of the $i$-th and $i'$-th nucleotides of the CGENM in these same directions are obtained by $CF^b_i=<|{\bf \delta x}_{i}{\bf b}_i|^2>$ and $CF^b_{i'}=<|{\bf \delta x}_{i'}{\bf b}_{i}|^2>$, $CF^s_{i}=<|{\bf \delta x}_{i}{\bf s}_{i}|^2>$ and $CF^s_{i'}=<|{\bf \delta x}_{i'}{\bf s}_i|^2>$, and $CF^t_{i}=<|{\bf \delta x}_{i}{\bf t}_i|^2>$ and $CF^t_{i'}=<|{\bf \delta x}_{i'}{\bf t}_i|^2>$, respectively.
Moreover, we consider the mean square fluctuation of the relative base position of each base pair of the CGENM in the three directions listed above given by $DF^b_i =<|({\bf \delta x}_{i}-{\bf \delta x}_{i'}){\bf b}_i|^2>$, $DF^s_i = <|({\bf \delta x}_{i}-{\bf \delta x}_{i'}){\bf s}_i|^2>$, and $DF^t_i = <|({\bf \delta x}_{i} - {\bf \delta x}_{i'}){\bf t}_i|^2>$, respectively.

\subsection{Evaluations of the Overall Geometry of DNA}
The overall geometry of each modeled DNA molecule is characterized by the ratios among the square root of the three principal components of the populations of atoms, $\sqrt{\lambda_1}$, $\sqrt{\lambda_2}$, and $\sqrt{\lambda_3}$ ($\lambda_1 > \lambda_2 > \lambda_3 > 0$). Here, $\lambda_1$, $\lambda_2$, and $\lambda_3$ are obtained as eigenvalues of the covariant matrix
\begin{equation}
I =
\begin{pmatrix}
<(\Delta x_i)^2>_i & <\Delta x_i \Delta y_i>_i & <\Delta x_i \Delta z_i>_i \\
<\Delta y_i \Delta x_i>_i & <(\Delta y_i)^2>_i & <\Delta y_i \Delta z_i>_i \\
<\Delta z_i \Delta x_i>_i & <\Delta z_i \Delta y_i>_i & <(\Delta z_i)^2>_i
\end{pmatrix},
\end{equation}
where $(\Delta x_i, \Delta y_i, \Delta z_i) = (x_i - x_{CM}, y_i - y_{CM}, z_i - z_{CM})$, $(x_i, y_i, z_i)$ is the position of $i$-th atom, $(x_{CM}, y_{CM}, z_{CM})$ is the position of the center of mass of a given DNA molecule, and $<...>_i$ indicates the average for all $i$s. We evaluate the overall geometry of a given DNA molecule using the linearity $\sigma_1 = \sqrt{\lambda_1}/\sqrt{\lambda_2}$ and the line symmetry with respect to the $\lambda_1$-axis $\sigma_2 = \sqrt{\lambda_3}/\sqrt{\lambda_2}$. Here, $\sigma_1$ is large when the DNA is straightened, and $\sigma_2$ is large (small) when the DNA forms a three (two)-dimensional curve with wide (flat) envelope.

\subsection{Evaluations of Correlations Among the Results of AAENM, CGENM, and Experiments}
We employ Pearson's correlation coefficients, $\rho$, to evaluate the correlations among the profiles of temperature factors and several anisotropic fluctuations of atoms obtained by AAENM, CGENM, and experiments.

\section{Results and Discussion}
\subsection{Comparisons of Fluctuations between the AAENM and the Crystal Structure of DNA}
The fluctuations of atoms of the AAENMs of several short DNA sequences were measured with normal-mode analysis. Here, the basic structures of the DNAs are given by the crystal structures of the naked DNAs with the following PDB IDs: 1BNA, 7BNA, 9BNA, 1D91, 1DC0, 122D, 123D, 181D, and 330D (Table 1\cite{1BNA,7BNA,9BNA,1D91,1DC0,122D_123D,181D,330D}). To confirm the validities of the AAENMs, the correlations between the distribution profiles of the temperature factor of atoms ($TF_i$) and the average temperature factor of the motifs ($MATF_i$) in the crystal structures and those in the corresponding AAENMs were measured. In the following arguments, we employ the optimal values of $C_a$, $B_B$, and $B_w$ for each crystal structure (Table 1), which were manually found to maximize $\rho$ of $TF_i$ between the results of the crystal structure analysis and those of the AAENM.

By choosing the appropriate conditions for the atoms of the two edge base pairs and water molecules ($B_B$ and $B_W$) for each PDB ID, $TF_i$ of each AAENM exhibited a similar profile to that of the crystal structure with a significant correlation coefficient $\rho$ (Fig. 2(a), Table 1, and \nameref{S1_Fig}). Therefore, the AAENM seemed to reproduce the overall structure of the temperature factor profile of each crystal structure  well, although the detailed profiles among atoms showed some deviations.

Furthermore, we focused on the average temperature factor for the motifs, $MATF_i$, as recently discussed \cite{full2}. We found that the $MATF_i$ obtained with the AAENMs could accurately reproduce those of the corresponding crystal structures with high correlation coefficients $\rho$, where $\rho > \sim 0.7$ was obtained for most cases (Fig. 2(b), Table 1, and \nameref{S2_Fig}). These results demonstrate that the AAENM is a suitable model for describing the sequence-dependent fluctuations of the nucleotide motifs of several double-stranded DNA sequences, despite the simplicity of model construction and its easy implementation. Moreover, these results show that the sequence-dependent forms of DNA have a dominant contribution to their overall flexibilities and fluctuations.

Nevertheless, the present AAENM could not accurately reproduce $MATF_i$ well for some crystal structures. These deviations are considered to have arisen from the following primary assumption: we only considered the effects of the restrictions by the packing of DNAs in crystal form for two edge base pairs and water, whereas such effects have an influence on all atoms. Therefore,  obtaining and incorporating more detailed knowledge of the restrictions of bulk sequences caused by crystal packing should help to achieve a more accurate reproduction of the molecular fluctuations for all cases.

\begin{figure}
\begin{center}
\includegraphics[width=8.0cm]{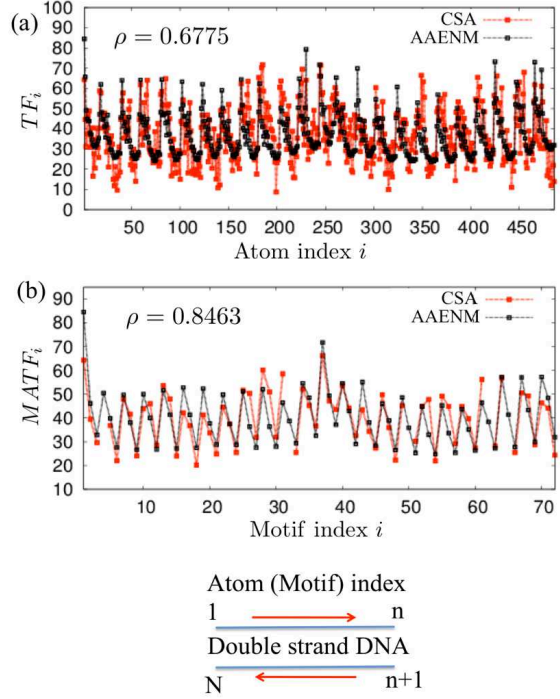}
\end{center}
\caption{{\bf Temperature factors obtained with the AAENMs and crystal structure analysis.} (a) Temperature factor of each atom ($TF_i$) and (b) average temperature factor of the motifs ($MATF_i$) obtained by the AAENMs (black curve) and crystal structure analysis (CSA, gray (red) curve) of typical double-stranded DNA (PDB ID: 1BNA). Here, $C_a =2.91 [kJ/(\AA^2 mol)]$, $B_B=0.018$, and $B_W=0.02$. Atom and motif indices in (a) and (b) are given in the same order as shown for (c). $\rho$ indicates the Pearson correlation coefficient of the profiles between the two curves.}
\end{figure}

\subsection{Comparison between the AAENMs and CGENMs of DNA}
The main objective of this study was to unveil the sequence-dependent dynamic correlated motions of several long DNA sequences. We next describe these dynamics of DNA sequences with longer lengths than considered in the previous subsection. For this purpose, we also constructed a coarse-grained model, which is often useful for focusing on the slow and large-scale dynamics of molecules that essentially influence their function. Thus, to propose a coarse-grained model of long double-stranded DNA, we evaluated whether the CGENM proposed provides an appropriate coarse-grained model of the present AAENM.

We performed normal-mode analysis of the AAENM and corresponding CGENM for $500$ randomly chosen $50$-bp sequences, and compared the mean square fluctuations of the $i$-th nucleotide ($NF_i$ and $CF_i$) in the directions parallel to the base pair axis ($NF_i^b$ and $CF_i^b$), parallel to the helix axis ($NF_i^s$ and $CF_i^s$), and in the torsional direction ($NF_i^t$ and $CF_i^t$). Here,  $C_a = 1.29 kJ/(\AA^2 mol)$ is employed, as in the previous study\cite{EN1}, and $C_g = 7.7 kJ/(\AA^2 mol)$ is assumed, which was manually found to provide the best fit of fluctuation profiles between AAENM and CGENM. Here, the overall fluctuation profiles of CGENM are independent of the value of $C_g$ since $C_g$ influences only on the absolute values of fluctuations. Independent of the sequences and employed helical parameters, we found that the fluctuations of each nucleotide in several directions were very similar between the AAENMs and CGENMs when these models are constructed with the same helical parameters, with average correlation coefficients $\rho$ $> 0.98$ (Fig. 3, Table 3, \nameref{S2_Table}, \nameref{S3_Fig}, and \nameref{S4_Fig}).

\begin{figure}
\begin{center}
\includegraphics[width=6.0cm]{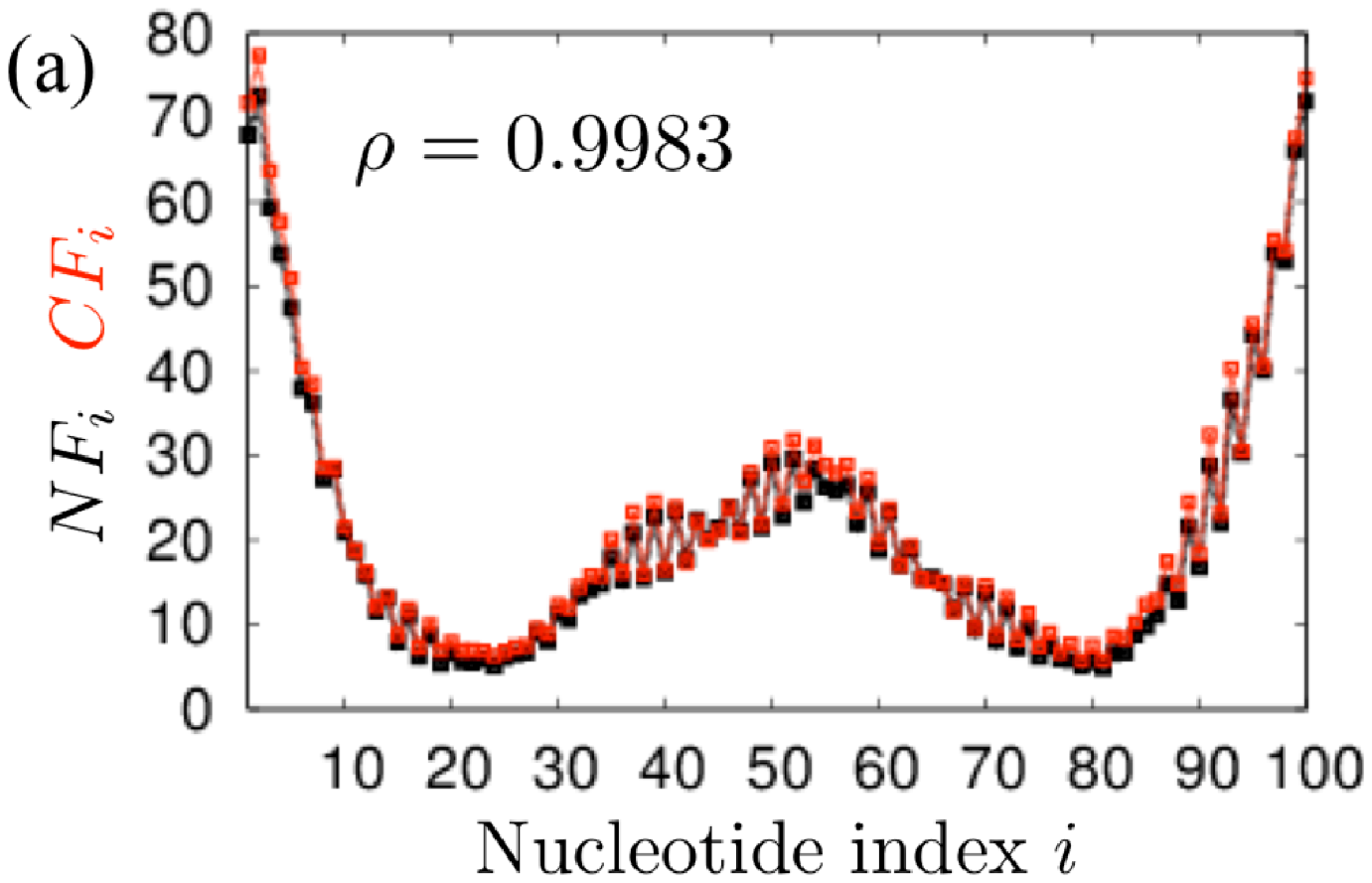}
\includegraphics[width=6.0cm]{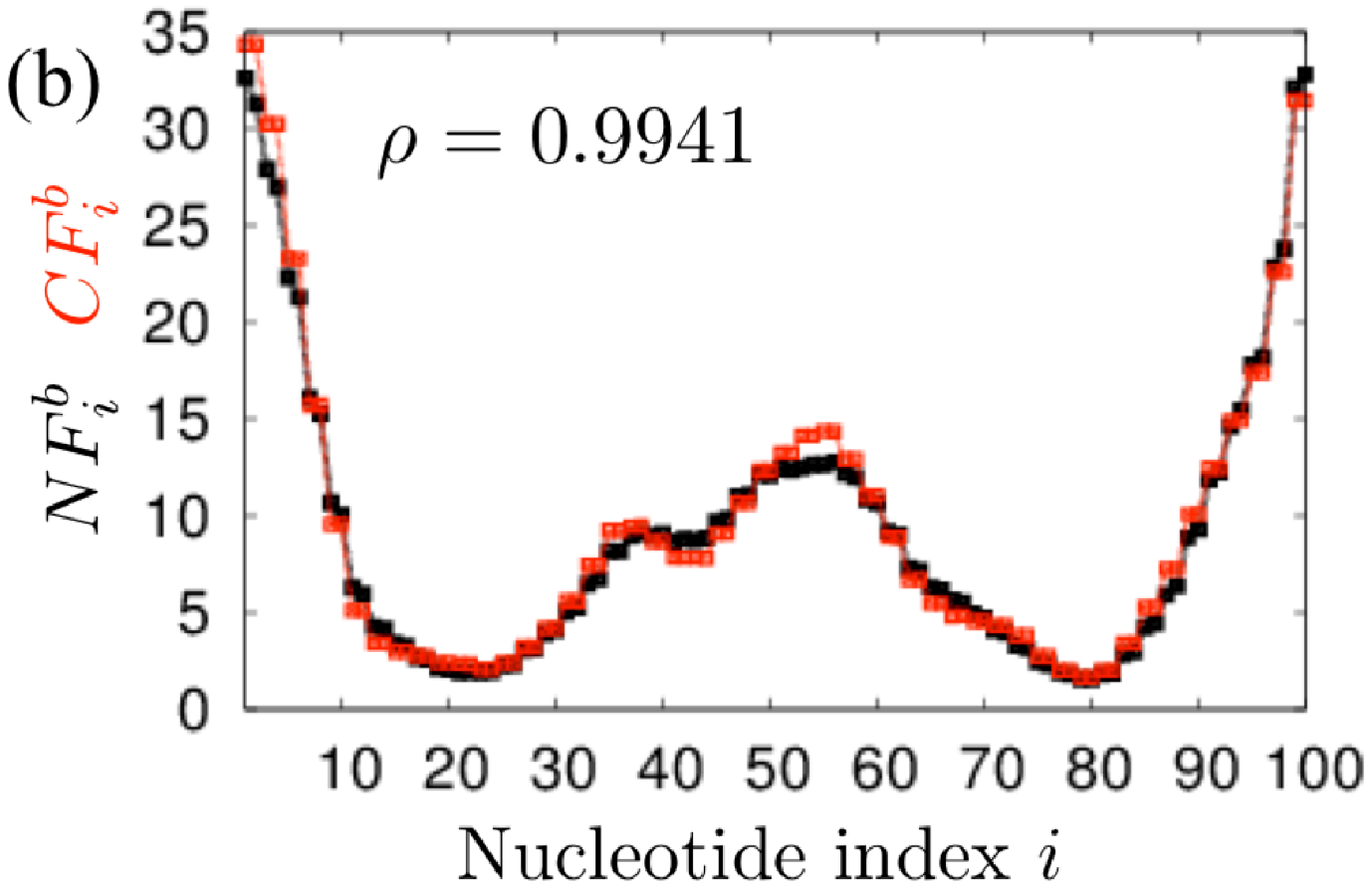}
\includegraphics[width=6.0cm]{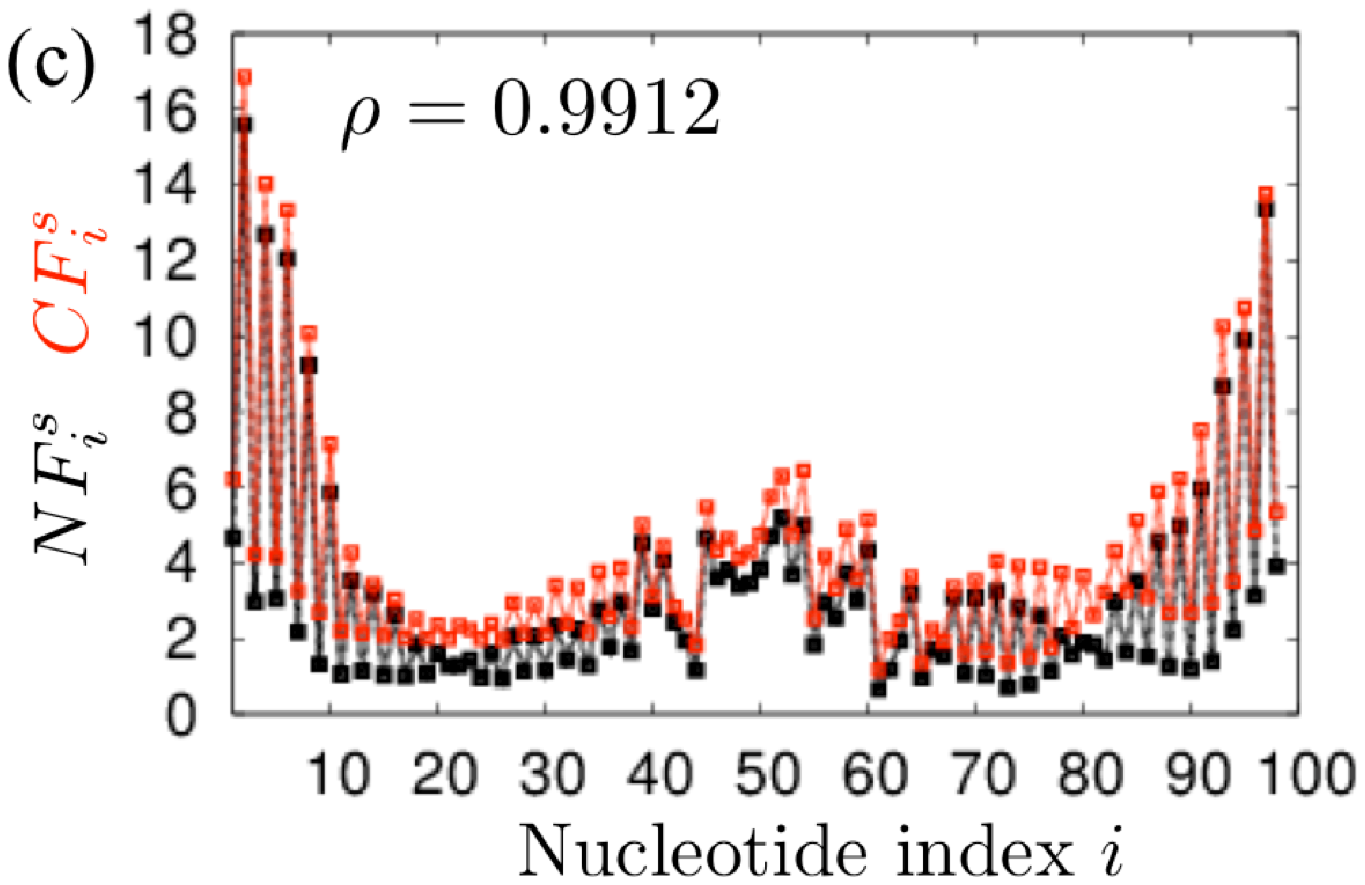}
\includegraphics[width=6.0cm]{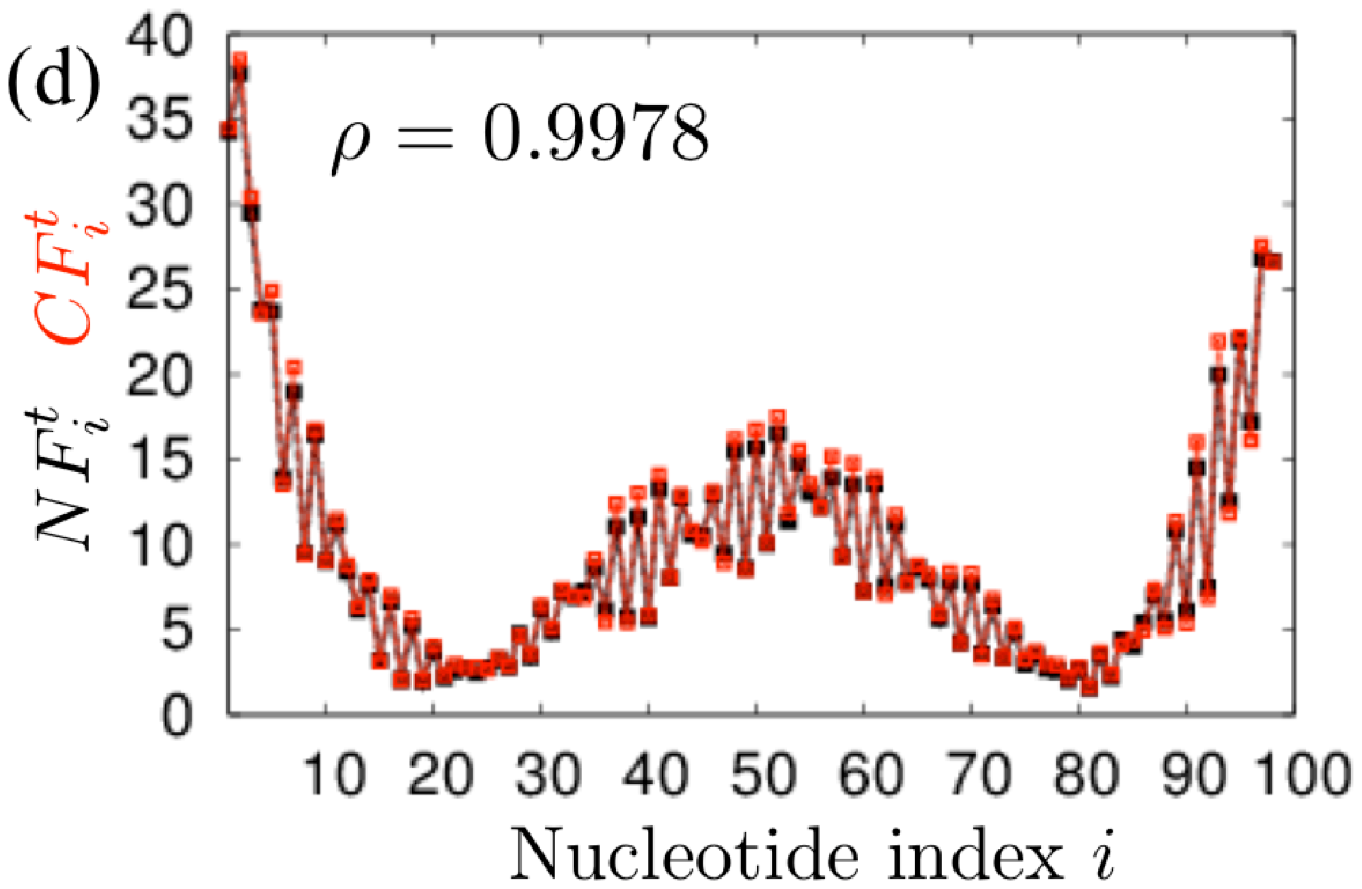}
\includegraphics[width=6.0cm]{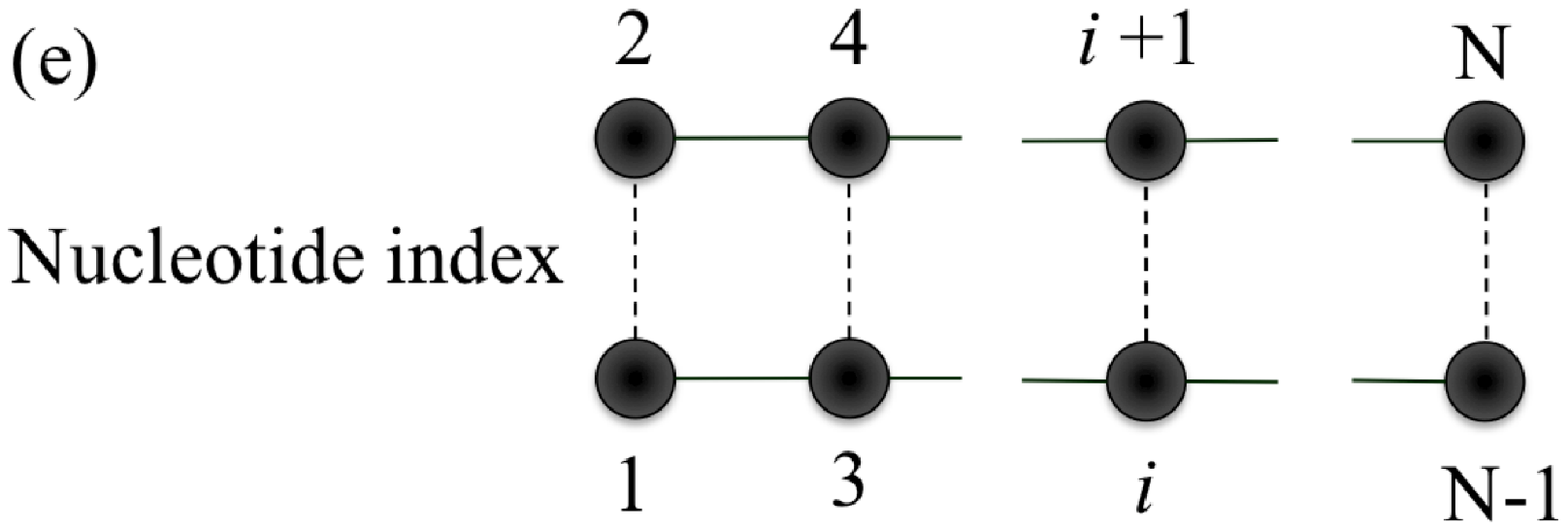}
\end{center}
\caption{{\bf Comparisons between the AAENM and CGENM.} Comparisons of the fluctuations between each nucleotide in the AAENM (black curves) and CGENM (gray (Red) curves) for a typical $50$-bp random DNA sequence (5' - GAGGCTAAAGTCTATTTAGACCGGAGTTGACGTGGAAGCCCGGCTAGTCT - 3'). (a) $NF_i$ and $CF_i$, (b) $NF_i^b$ and $CF_i^b$, (c) $NF_i^s$ and $CF_i^s$, and (d) $NF_i^t$ and $CF_i^t$. Helical parameter set (i) (Table 2) was used for both models. The nucleotide indices in (a) to (d) are given in the same order shown for (e). $\rho$ indicates the Pearson correlation coefficient of the profiles between the two curves.}
\end{figure}

It is noted that the present CGENM contains only one node per nucleotide, whereas the AAENM contains $19 \sim 22$ nodes (atoms) per nucleotide. This fact demonstrates that the computational costs of the CGENMs are much lower than those of the AAENMs, although the accuracies of the obtained statistical aspects are almost identical between these two models. Thus, this CGENM could be used for exhaustive analysis and comparisons of the dynamic features of several sequence-dependent DNAs related to protein binding affinities, functions of transcription regulation sequences, and nucleosome positioning\cite{gene0,gene1,gene2,gene3,gene4,ins1,ins2,ins3,ins4,ins5,ins6,cg4,nuc1,nuc2,nuc3,nuc4,nuc5,nuc6}. In the next subsection, we provide an example of such an analysis to determine the relationships between the nucleosome-forming abilities of several double-stranded DNA sequences and their inter-strand dynamic features.

\begin{table}
\begin{center}
\includegraphics[width=6.0cm]{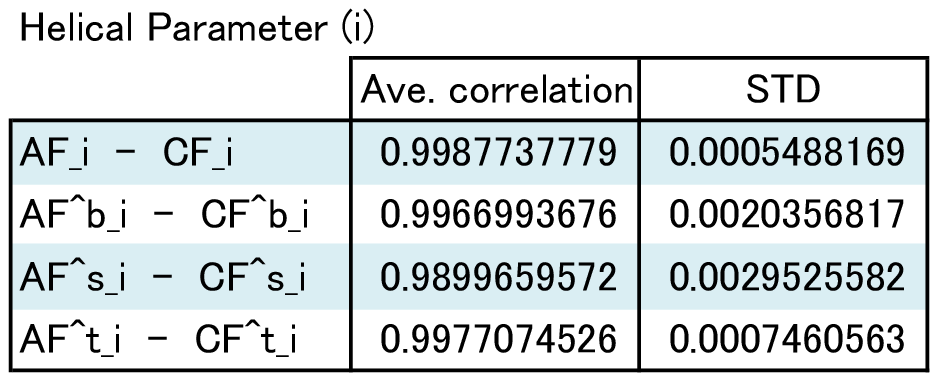}
\end{center}
\caption{{\bf Comparisons between the AAENM and CGENM.} Average and standard deviation of the correlation coefficients of 500 random samples of 50-bp sequences between $NF_i$ and $CF_i$, $NF_i^b$ and $CF_i^b$, $NF_i^s$ and $CF_i^s$, and $NF_i^t$ and $CF_i^t$. Helical parameter set (i) (Table 2) was used in all cases. }
\end{table}

\subsection{Exhaustive Analysis of the Sequence-dependent Behaviors of $150$-bp DNAs with the CGENM}
Nucleosome positioning is important not only for compacting DNA but also for appropriate gene regulation. Several recent studies have been performed for genome-wide nucleosome mapping and the identifications and predictions of nucleosome-forming and -inhibiting sequences for some model organisms\cite{nuc3,nuc5,yeastnuc0,yeastnuc1,homosapi,hae1,hae2}.

As an example of the applications of the CGENM to an exhaustive analysis of the sequence-dependent behavior of DNA, we compared the dynamic features of DNA sequences of $\sim 150$ bp that were predicted as nucleosome-forming or nucleosome-inhibiting sequences in the genome of budding yeast {\it Saccharomyces cerevisiae} (5000 forming sequences and 5000 inhibiting sequences of $150$ bp)\cite{nuc3}, nematode {\it Caenorhabditis elegans} (2567 forming sequences and 2608 inhibiting sequences of $147$ bp), {\it Drosophila melanogaster} (2900 forming sequences and 2850 inhibiting sequences of $147$ bp), and {\it Homo sapiens} (2273 forming sequences and 2300 inhibiting sequences of $147$ bp)\cite{nuc5}. The histograms of the average relative fluctuations of DNAs for the three directions $<CF_i>_i$, $<CF^b_i>_i$, $<CF^s_i>_i$, $<CF^t_i>_i$, $<DF_i>_i$, $<DF_i^b>_i$, $<DF_i^s>_i$, and $<DF_i^t>_i$ ($<...>_i$ indicates the average for all $i$s.) for the nucleosome-forming sequences and the nucleosome-inhibiting sequences are shown in Fig. 4 and \nameref{S5_Fig} $\sim$ \nameref{S7_Fig}. Here, we employed the helical parameter set (i) (Table 2) that was used in the coarse-grained molecular dynamics simulations by Freeman et al., which exhibited consistent results to some experiments\cite{cg6,heri1,heri2,heri3}.

The histograms for budding yeast showed that that nucleosome-forming sequences tend to exhibit larger fluctuations in several directions compared to the inhibiting sequences. In particular, the histogram of $<DF_i^b>_i$ for the nucleosome-forming sequences showed a clear shift in the direction toward larger values compared to that for the nucleosome-inhibiting sequences (Fig. 4). For the other organisms, most of the histograms showed few differences between the nucleosome- forming and -inhibiting sequences. However, similar to the case of yeast, the distribution of $<DF_i^b>_i$ for the nucleosome-forming sequences shifted largely in the direction of larger values compared to that for the nucleosome-inhibiting sequences in these organisms (\nameref{S5_Fig} $\sim$ \nameref{S7_Fig}). These results indicate that the nucleosome-forming ability is highly correlated to the fluctuations of the inter-strand distances of DNAs, in which sequences with larger fluctuations tend to form the nucleosome.

\begin{figure}
\begin{center}
\includegraphics[width=12.0cm]{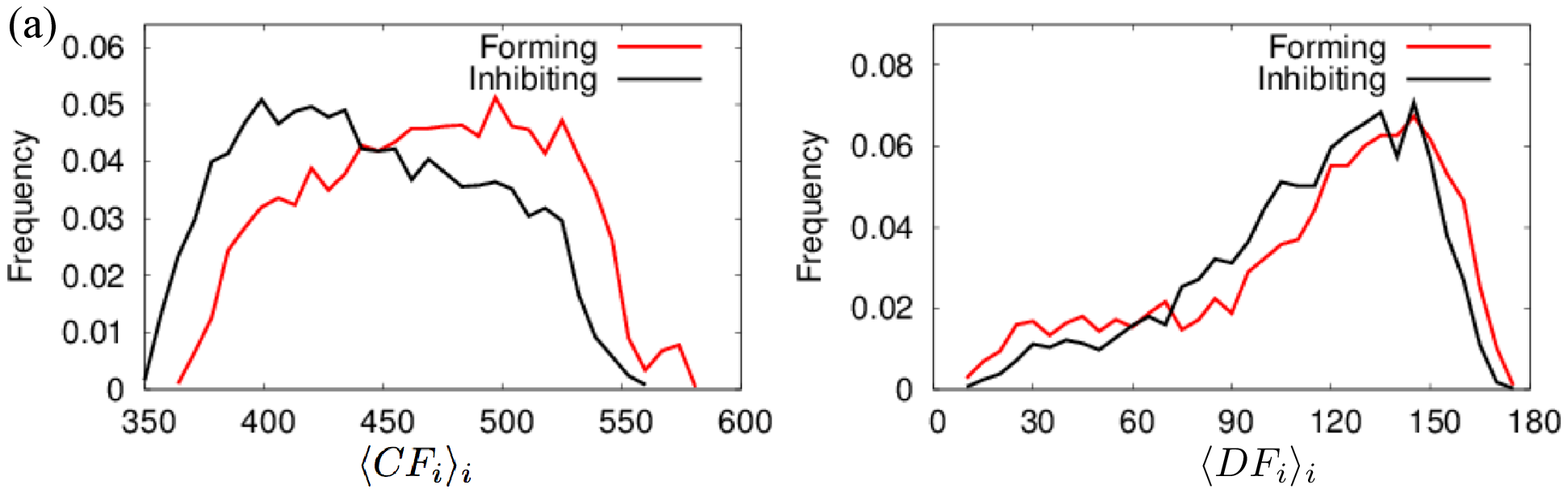}
\includegraphics[width=12.0cm]{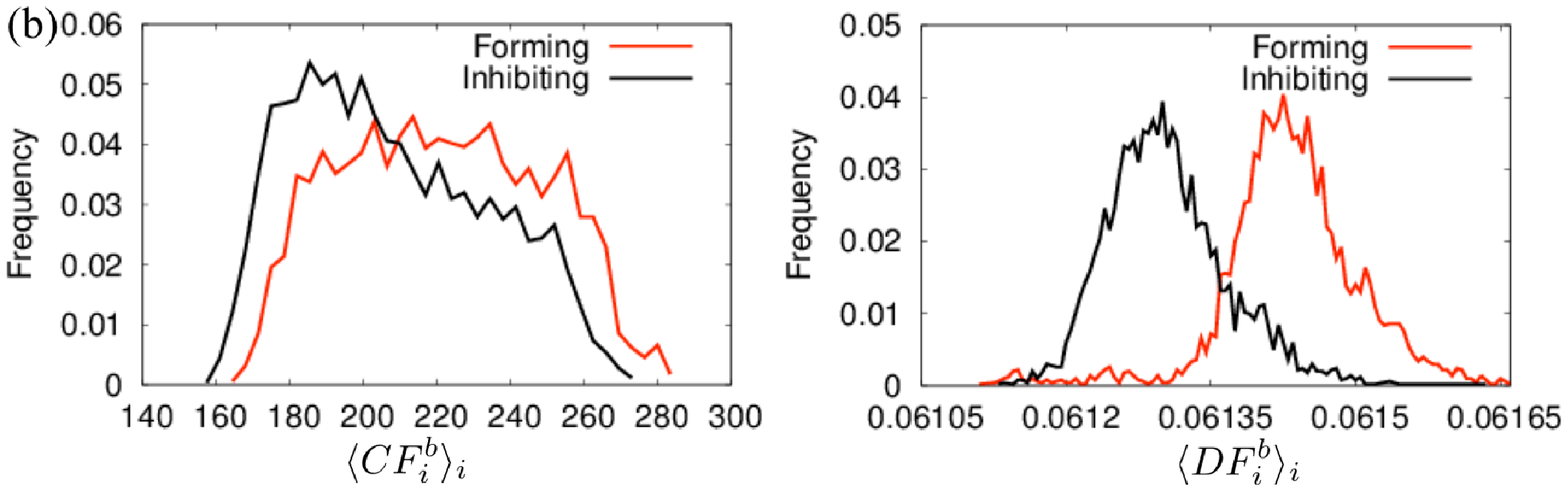}
\includegraphics[width=12.0cm]{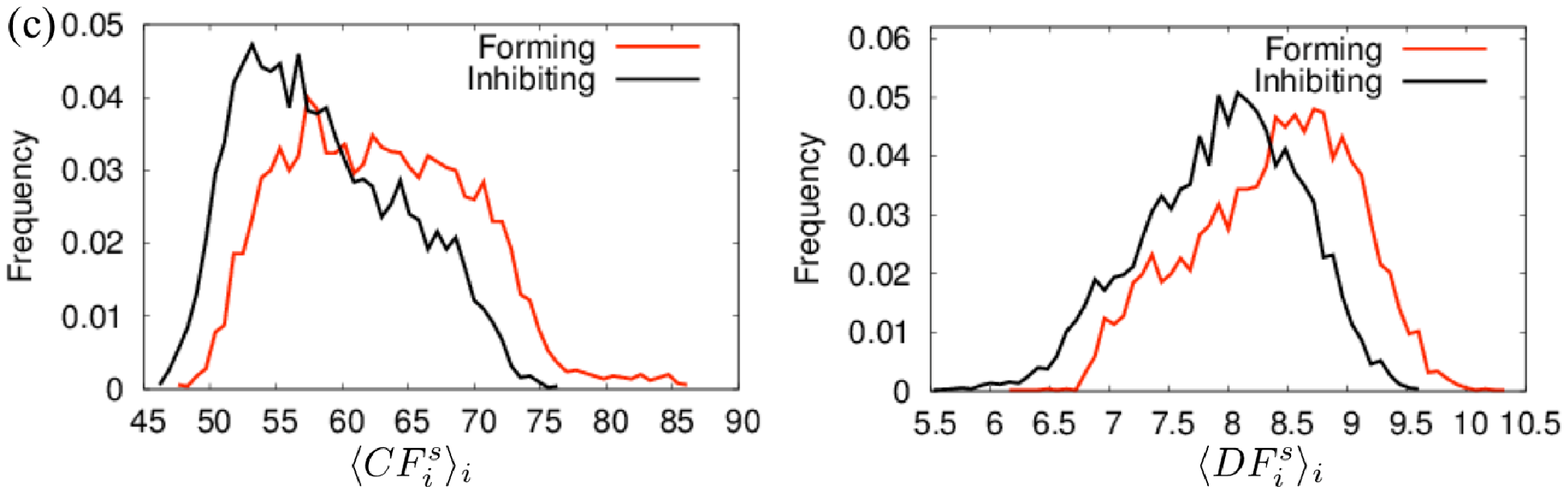}
\includegraphics[width=12.0cm]{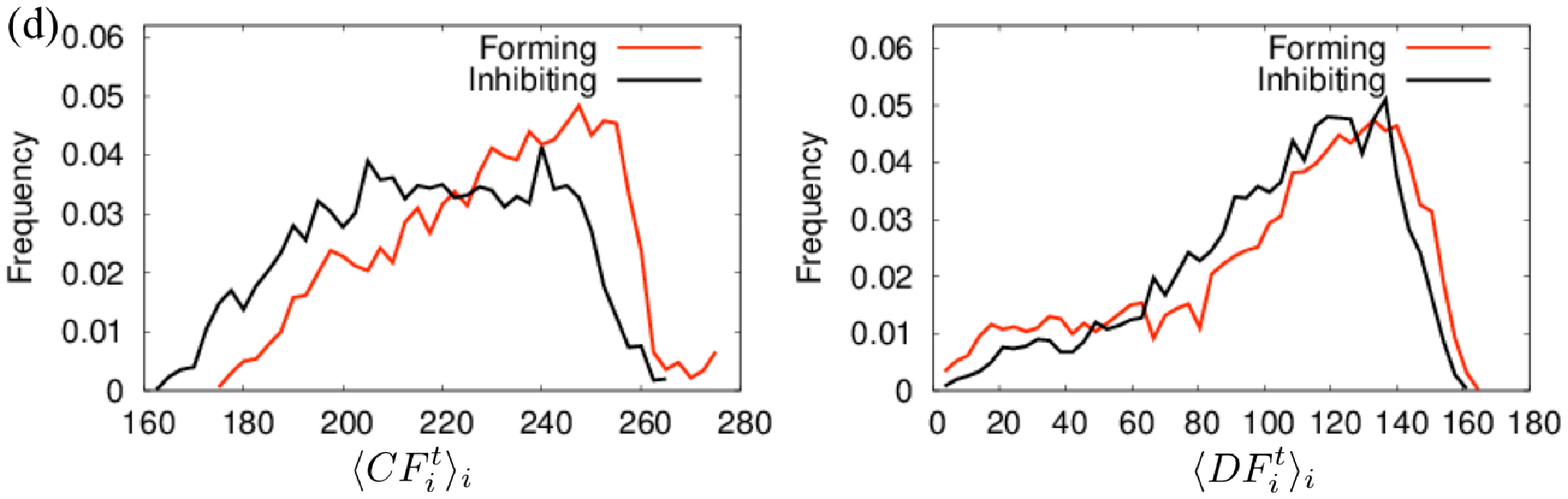}
\end{center}
\caption{{\bf Fluctuations of the CGENM of long DNA sequences.} Histograms of the average fluctuations, (a) $<CF_i>_i$ and $<DF_i>_i$, (b) $<CF_i^b>_i$ and $<DF_i^b>_i$, (c) $<CF_i^s>_i$ and $<DF_i^s>_i$, and (d) $<CF_i^t>_i$, and $<DF_i^t>_i$, for nucleosome-forming and -inhibiting sequences of budding yeast{\it Saccharomyces cerevisiae} ($150$ bp). Helical parameter set (i) (Table 2) was used.}
\end{figure}

Similar to the previous arguments, we compared the dynamical features among the random $150$-bp  DNA sequences varying in average GC-contents; the histograms, averages, and standard deviations of $<CF_i>_i$, $<DF_i>_i$, $<DF_i^b>_i$, $<DF_i^s>_i$, and $<DF_i^t>_i$ were measured from 10,000 random sequences for each GC content (Fig. 5 and \nameref{S8_Fig}). In this case, $<CF_i>_i$ exhibited a minimum at a GC content of $\sim 0.2$, which indicates that the AT-rich sequences tend to be more rigid than the GC-rich sequences. However, the fluctuations of sequences consisting of only A or T were as large as those of the GC-rich sequences. Moreover, the GC content dependencies of $<DF_i>_i$, $<DF_i^b>_i$, $<DF_i^s>_i$, and $<DF_i^t>_i$ showed different characteristics for GC contents larger or smaller than $0.6 \sim 0.7$. In particular, the results for $<DF_i^b>_i$ were similar for cases with a large GC content ($0.7 \sim 1$) but monotonically decreased with a GC content with little variance. A recent experimental study showed that the probability of  nucleosome formation tends to increase with increases in the GC content ratio \cite{gcnuc}. Thus, the present results indicate that sequences with larger fluctuations of inter-strand distances tend to form the nucleosome, which is consistent with the results described above from the analysis of the genomes of the four model organisms.

\begin{figure}
\begin{center}
\includegraphics[width=10.0cm]{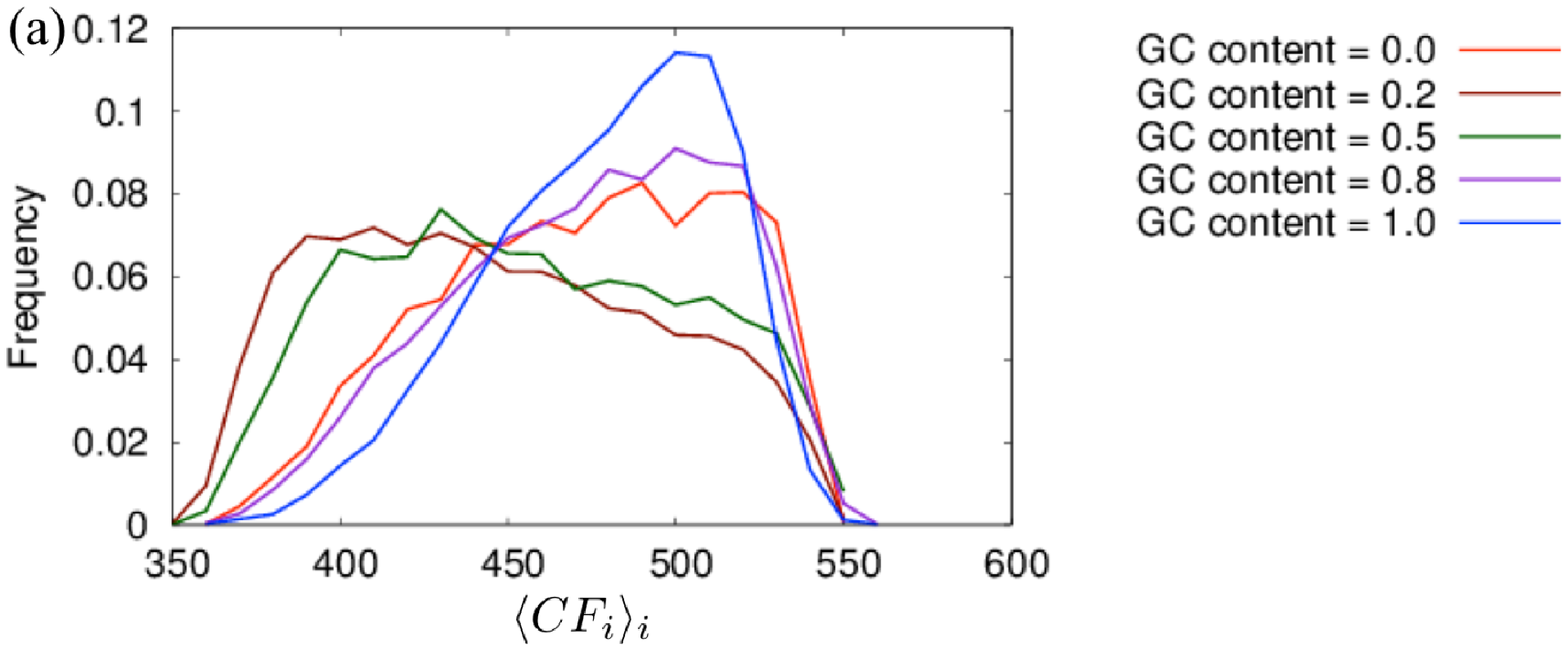}
\includegraphics[width=10.0cm]{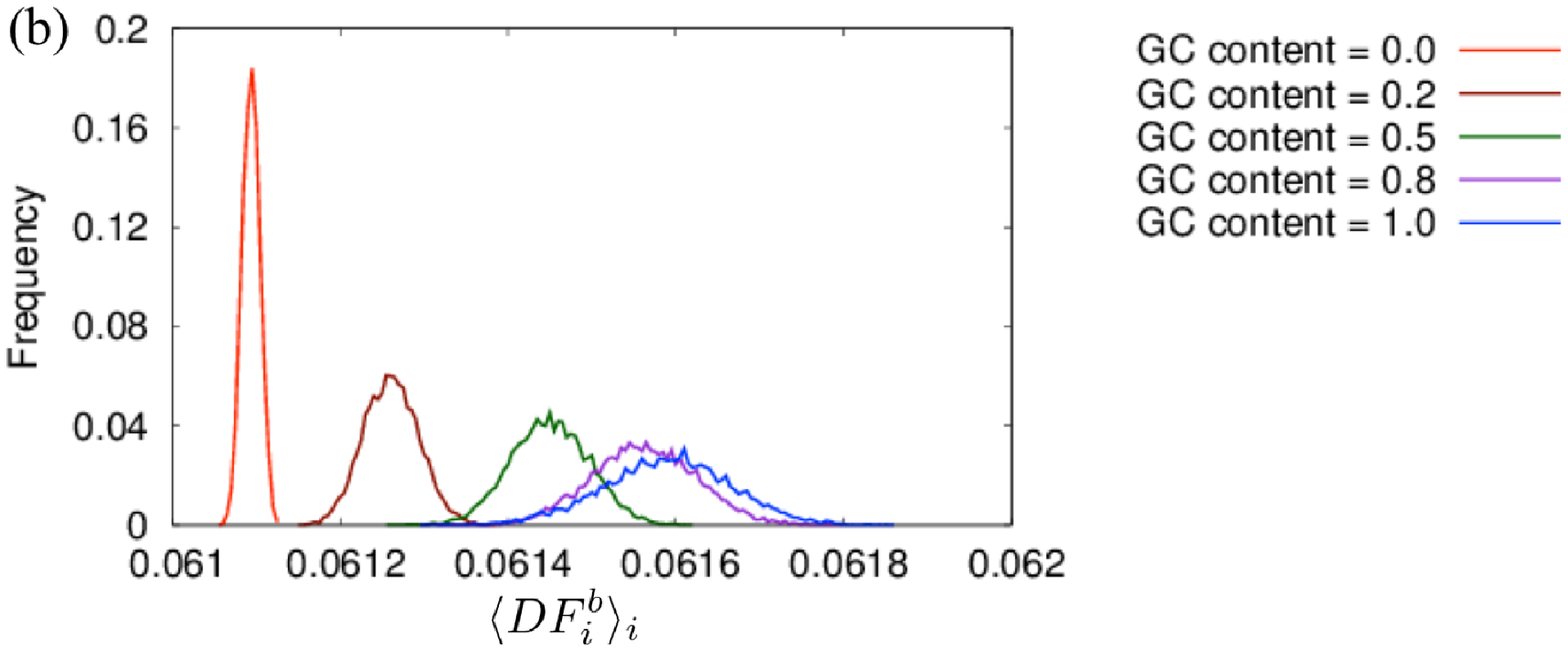}
\includegraphics[width=12.0cm]{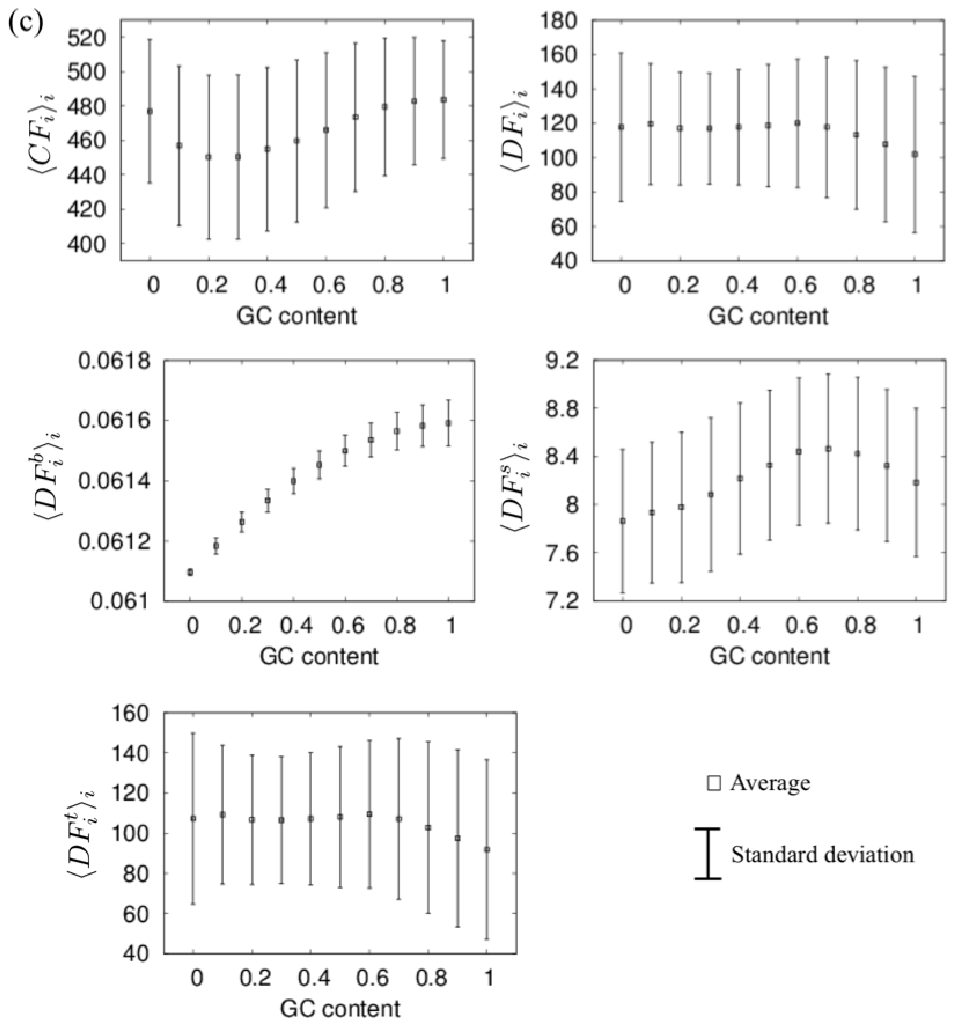}
\end{center}
\caption{{\bf Fluctuations of the CGENM of long DNA sequences.} Histograms of (a) $<CF_i>_i$ and (b) $<DF_i^b>_i$, and (c) averages and standard deviations of $<CF_i>_i$, $<DF_i>_i$, $<DF_i^b>_i$, $<DF_i^s>_i$, and $<DF_i^t>_i$ for 10,000 samples of random $150$-bp sequences with an average GC content $= 0$, $0.1$, $0.2$, $\cdots$, and $1$. Helical parameter set (i) (Table 2) was used.}
\end{figure}

Finally, we focus on the relationships between the overall geometries and fluctuations of the considered DNA sequences. The overall geometries of several DNA sequences, the nucleosome-forming and -inhibiting DNA sequences for four model organisms and random DNA sequences with different GC contents, were evaluated using scatter plots of $\sigma_1$ (linearity) and $\sigma_2$ (line symmetry) (Fig. 6 and \nameref{S9_Fig}). It is noted that $\sigma_1$ and $\sigma_2$ are highly correlated. The average, standard deviation, and distributions of $\sigma_1$ and $\sigma_2$ exhibited slight but not significant deviations between the nucleosome-forming and -inhibiting sequences.

For the random sequences, the average value of $\sigma_1$ exhibited similar variations to $<CF_i>_i$ with an increase in GC content. In particular, both values decreased with increases in GC content for GC contents $\le 0.2$, whereas they increased with increases in GC content for GC contents $\ge 0.3$ (Fig. 5(c) and Fig. 6(c)). In fact, $\sigma_1$ and $<CF_i>$ were highly correlated for random DNA sequences, regardless of their CG content. The Pearson correlation coefficient for the 110,000 sequences analyzed above with GC contents $= 0.0 \sim 1.0$ was $0.9433$. It is noted that $\sigma_1$ showed large dispersion for each GC content, and there were significant overlaps among $\sigma_1$ distributions with different GC contents (Fig 6(b) and 6(c)). This fact indicates that different DNA sequences can often show similar geometries, and such sequences also tend to show similar overall fluctuations. On the other hand, the fluctuations of inter-strand distances $<DF^b_i>$ that may correlate to the nucleosome-forming ability did not correlate significantly to either $\sigma_1$ or $\sigma_2$, with Pearson correlation coefficients of $0.2250 $ and $0.1967$. This fact indicates that the nucleosome-forming ability of DNA sequences are not only determined by the overall DNA geometries but also by their dynamic properties.

\begin{figure}
\begin{center}
\includegraphics[width=12.0cm]{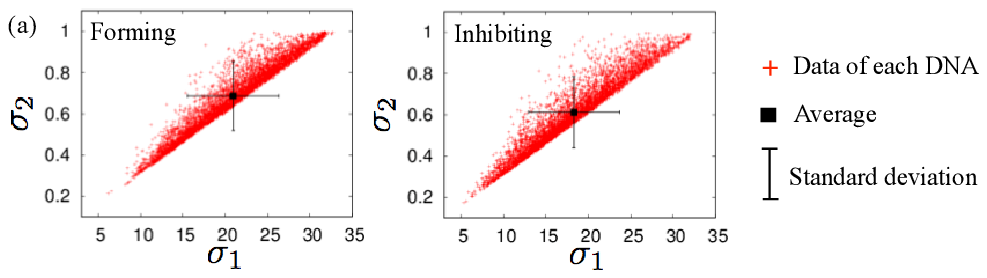}
\includegraphics[width=12.0cm]{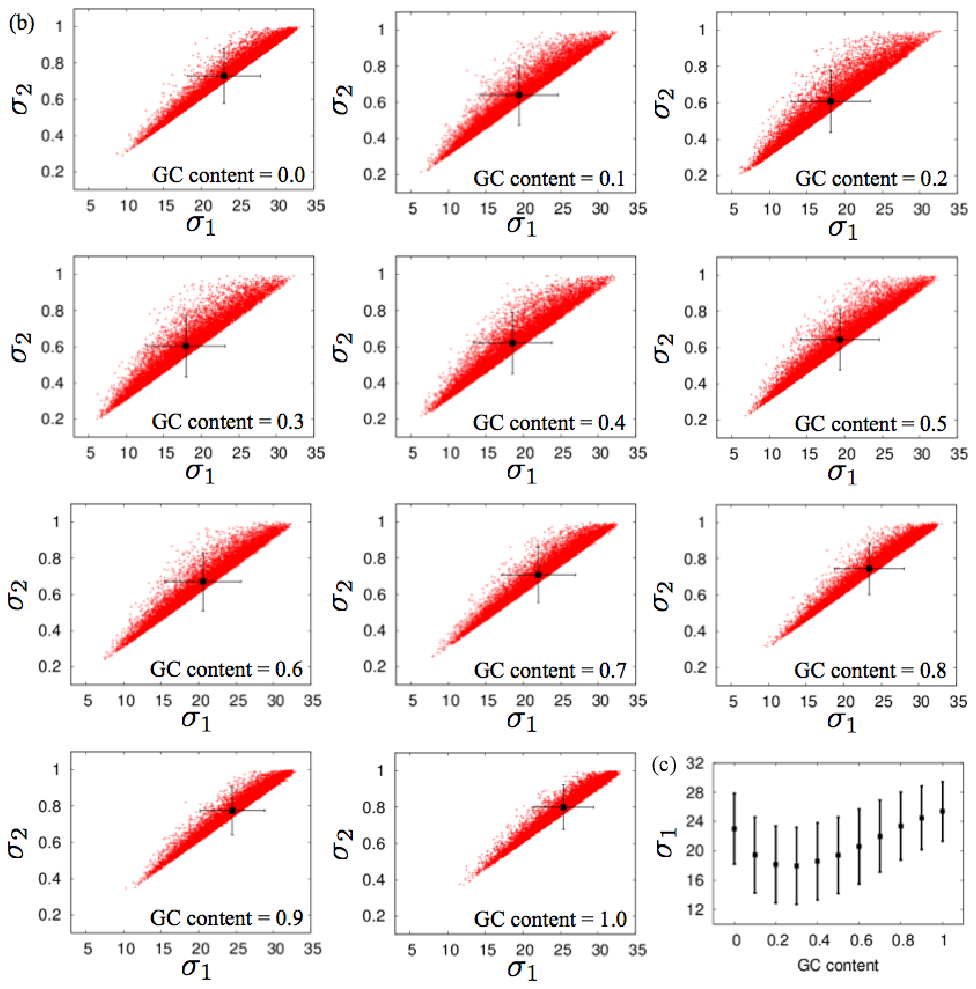}
\end{center}
\caption{{\bf Overall geometries of long DNA sequences.} Scatter plots of $\sigma_1$ and $\sigma_2$ for (a) nucleosome-forming and -inhibiting sequences of budding yeast{\it Saccharomyces cerevisiae} ($150$ bp), and for (b) random sequences with several GC contents ($150$ bp). (c) Averages and standard deviations of $\sigma_1$ for 10000 samples of random $150$-bp sequences with an average GC content $= 0$, $0.1$, $0.2$, $\cdots$, and $1$. Helical parameter set (i) (Table 2) was used.}
\end{figure}

\section{Summary and Conclusion}
In this study, simple elastic network models of double-stranded DNAs were developed in order to perform an exhaustive analysis of several sequence-dependent dynamical features. First, we constructed a simple all-atom elastic network model that could reproduce the fluctuations of the motifs of each nucleotide (sugar, phosphoric acid, and bases) of several crystal structures of short DNA sequences. Second, we proposed a simple course-grained elastic network model that could reproduce the dynamic features of the long DNA sequences obtained by the all-atom elastic network model. Through exhaustive analysis of the dynamic features of several DNA sequences with normal-mode analysis of the presented coarse-grained elastic network model, we found that the dynamic aspects of DNA are highly influenced by the properties of nucleotide sequence such as GC content. We also found that the nucleosome-forming abilities of double-stranded DNA exhibited positive correlations with their sequence-dependent inter-strand fluctuations.

In the present study, we demonstrated the sequence-dependent dynamic features for several $\sim 150$-bp DNA sequences to evaluate the relationships between the nucleosome-forming abilities and DNA dynamics. Of course, DNA sequences longer than $\sim 150$ bp can also be analyzed using the presented course-grained model. Moreover, course-grained molecular dynamics simulations can be performed to consider the large deformations of DNA, such as formation of a super helix and nucleosome that are the basic structures of higher-order chromosome architectures, using the presented elastic network models with the excluded effect of the volume of each atom or each nucleotide. We are currently conducting these molecular dynamics simulations, and the results will be reported in the future. We did not consider the effects of solvents such as temperature and salt concentrations in the present elastic network models. Therefore, we are also planning to attempt modifications of the models so that several solvent conditions can be incorporated in future work.

\section*{Acknowledgments}
The author is grateful to S. Tate and Y. Murayama for fruitful discussions.
\\[3ex]


\section*{Author contributions}
Conceived the study: AA, NS, HN. Designed the models: NS, AA, SI. Performed the analysis: SI, AA. Wrote the paper: AA, NS, HN.


\section*{Supporting Information}

\subsection*{S1 Table}
\label{S1_Table}

{\bf Helical parameter sets.} (a) Helical parameter sets (ii) obtained by X-ray crystal structure analysis,\cite{heri1,heri2}, and (b) helical parameter sets (iii) obtained by all-atom molecular dynamics simulations.\cite{nuc4,heri5,heri6}.

\subsection*{S2 Table}
\label{S2_Table}

{\bf Comparisons between CGENM and AAENM.} Average and standard deviation of the correlation coefficients of 500 samples of random $50$-bp sequences between $NF_i$ and $CF_i$, $NF_i^b$ and $CF_i^b$, $NF_i^s$, and $CF_i^s$, and $NF_i^t$ and $CF_i^t$. Helical parameter sets (ii) and (iii) (\nameref{S1_Table}) were used.

\subsection*{S1 Fig}
\label{S1_Fig}

{\bf Temperature factor of each atom.} The temperature factor of each atom obtained by the AAENMs (black curve) and X-ray crystal structure analysis (gray (red) curve) of typical double-stranded DNAs obtained from PDB ID (a) 7BNA, (b) 9BNA, (c) 1D91, (d) 1DC0, (e) 122D, (f) 123D, (g) 181D, and (h) 330D. Parameters $C_a$, $B_B$, and $B_W$ are given in \nameref{S1_Table}.

\subsection*{S2 Fig}
\label{S2_Fig}

{\bf Average temperature factor of motifs.} Average temperature factor of motifs obtained by the AAENMs (black curve) and X-ray crystal structure analysis (gray (red) curve) of typical double-stranded DNAs obtained from PDB ID (a) 7BNA, (b) 9BNA, (c) 1D91, (d) 1DC0, (e) 122D, (f) 123D, (g) 181D, and (h) 330D. Parameters $C_a$, $B_B$, and $B_W$ are given in \nameref{S1_Table}.

\subsection*{S3 Fig}
\label{S3_Fig}

{\bf Comparisons between AAENM and CGENM.} Comparisons of the fluctuations between each nucleotide in the AAENM (black curves) and CGENM (gray (red) curves) for a typical random $50$-bp DNA sequence (5' - AGTGGTAAGGCATGGTTCTCGAATCTCGGTTTATTTACACTGCTGCTCCA - 3'). (a) $NF_i$ and $CF_i$, (b) $NF_i^b$ and $CF_i^b$, (c) $NF_i^s$ and $CF_i^s$, and (d) $NF_i^t$ and $CF_i^t$ using helical parameter set (ii) \nameref{S2_Table}.

\subsection*{S4 Fig}
\label{S4_Fig}

{\bf Comparisons between AAENM and CGENM.} Comparisons of the fluctuations between each nucleotide in the AAENM (black curves) and CGENM (gray (red) curves) for a typical random $50$-bp DNA sequence (5' - ATATGCTGTAGAGCGTCCCGTCCGCGCGTTGTGGTTTTTTCGGTGCTCTA - 3'). (a) $NF_i$ and $CF_i$, (b) $NF_i^b$ and $CF_i^b$, (c) $NF_i^s$ and $CF_i^s$, and (d) $NF_i^t$ and $CF_i^t$ using helical parameter set (iii) \nameref{S1_Table}.

\subsection*{S5 Fig}
\label{S5_Fig}

{\bf Histograms of the average fluctuations in{\it Caenorhabditis elegans}.} Histograms of the average fluctuations of (a) $<CF_i>_i$ and $<DF_i>_i$, (b) $<CF_i^b>_i$ and $<DF_i^b>_i$, (c) $<CF_i^s>_i$ and $<DF_i^s>_i$, and (d) $<CF_i^t>_i$ and $<DF_i^t>_i$ for nucleosome-forming and nucleosome-inhibiting sequences of the nematode {\it Caenorhabditis elegans} ($147$ bp). Helical parameter set (i) (Table 2) was used.

\subsection*{S6 Fig}
\label{S6_Fig}

{\bf Histograms of the average fluctuations in{\it Drosophila melanogaster}.} Histograms of the average fluctuations of (a) $<CF_i>_i$ and $<DF_i>_i$, (b) $<CF_i^b>_i$ and $<DF_i^b>_i$, (c) $<CF_i^s>_i$ and $<DF_i^s>_i$, and (d) $<CF_i^t>_i$ and $<DF_i^t>_i$ for nucleosome-forming and nucleosome-inhibiting sequences of {\it Drosophila melanogaster} ($147$ bp). Helical parameter set (i) (Table 2) was used.

\subsection*{S7 Fig}
\label{S7_Fig}

{\bf Histograms of the average fluctuations in{\it Homo sapiens}.} Histograms of the average fluctuations of (a) $<CF_i>_i$ and $<DF_i>_i$, (b) $<CF_i^b>_i$ and $<DF_i^b>_i$, (c) $<CF_i^s>_i$ and $<DF_i^s>_i$, and (d) $<CF_i^t>_i$ and $<DF_i^t>_i$ for nucleosome-forming and nucleosome-inhibiting sequences of {\it Homo sapiens} ($147$ bp). Helical parameter set (i)  (Table 2) was used.

\subsection*{S8 Fig}
\label{S8_Fig}

{\bf Histograms of the average fluctuations with different GC contents.} Histograms of (a) $<DF_i>_i$, (b) $<DF^s_i>_i$, and (c) $<DF^t_i>_i$ for 10,000 samples of random $150$-bp sequences with different average GC contents. Helical parameter set (i) (Table 2) was used.

\subsection*{S9 Fig}
\label{S9_Fig}

{\bf Overall geometries of long DNA sequences in model organisms.} Scatter plots of $\sigma_1$ and $\sigma_2$ for nucleosome-forming and -inhibiting sequences of (a) {\it Caenorhabditis elegans} ($147$ bp), (b) {\it Drosophila melanogaster} ($147$ bp), and (c) {\it Homo sapiens} ($147$ bp). Helical parameter set (i) (Table 2) was used.

\end{document}